\documentstyle{article}
\begin{document}

\def\sec#1{\vspace*{1\baselineskip}\centerline{} \vspace*{0.5\baselineskip}}
\def\subsec#1{\vspace*{1\baselineskip} \leftline{}}
\renewcommand{\thefootnote}{\fnsymbol{footnote}}
\renewcommand{\theequation}{\arabic{equation}}
\newcommand{\beq}{\begin{equation}}
\newcommand{\eeq}{\end{equation}}
\newcommand{\diag}{\mbox{diag}}

\vspace*{1cm}
\begin{center}
{{\LARGE On the Number of Positive Solutions to a Class of Integral Equations
\LARGE $^{\ast }$
}}
\end{center}

\footnotetext
{\small $\ast$Supported by the National Natural Science Foundation of China
(No. 69925307), National Key Project of China,
National Key Basic Research Special Funds of China (No. G1998020302)
and National Laboratory of Intelligent Technology and Systems of Tsinghua University.
Corresponding author: Professor Long Wang, Email: longwang@mech.pku.edu.cn}

\vskip 0.8cm \centerline{Long Wang} \vskip 6pt \centerline{{Center
for Systems and Control, Department of Mechanics and Engineering
Science}} \centerline{ {Peking University, Beijing 100871, P. R.
CHINA}}

\vskip 0.5cm \centerline{Wensheng Yu} \vskip 6pt
\centerline{{Laboratory for Complex Systems and Intelligent
Control, Institute of Automation}} \centerline{ {Chinese Academy
of Sciences, Beijing 100080, P. R. CHINA}}

\vskip 0.5cm \centerline{Lin Zhang} \vskip 6pt \centerline{
{Department of Automation, Tsinghua University}} \centerline{
{Beijing 100084, P. R. CHINA}}

\vskip 0.6cm

{Abstract:}
\begin{minipage}[t]{11cm}{
By using the complete discrimination system for polynomials, we study the number of
positive solutions in {\small $C[0,1]$} to the integral equation
{\small $\varphi (x)=\int_0^1k(x,y)\varphi ^n(y)dy$}, where
{\small $k(x,y)=\varphi _1(x)\phi _1(y)+\varphi _2(x)\phi _2(y),
\varphi _i(x)>0, \phi _i(y)>0, 0<x,y<1, i=1,2,$}
are continuous functions on {\small $[0,1]$},
{\small $n$} is a positive integer. We prove the following results: when
{\small $n= 1$}, either there does not exist, or there exist infinitely
many positive solutions in {\small $C[0,1]$}; when {\small $n\geq 2$}, there exist
at least {\small $1$}, at most {\small $n+1$} positive solutions in {\small $C[0,1]$}.
Necessary and sufficient conditions are derived for the cases: 1) {\small $n= 1$}, there
exist positive solutions; 2) {\small $n\geq 2$},
there exist exactly {\small $m(m\in \{1,2,\cdots,n+1\})$}
positive solutions. Our results generalize the existing results in the literature, and
their usefulness is shown by examples presented in this paper.
}
\end{minipage}

\vskip 0.3cm
{Keywords:}
\begin{minipage}[t]{11cm}
{

Integral Equations, Positive Solutions, the Complete Discrimination System for Polynomials,
the Number of Solutions}
\end{minipage}

\vskip 0.8cm
\section{ Introduction}

\vskip 0.5cm
The existence of positive solutions to integral equations is an active research
field and has important applications in the stability of feedback systems [1,2].
In 1991, the number of positive solutions to the following integral
equation

\begin{equation}
\varphi (x)=\int_0^1k(x,y)\varphi ^2(y)dy
\end{equation}
was discussed in [3]. In this paper, we will study the number of positive solutions
in $C[0,1]$ to following more general integral equation

\begin{equation}
\varphi (x)=\int_0^1k(x,y)\varphi ^n(y)dy
\end{equation}
where $$k(x,y)=\varphi _1(x)\phi _1(y)+\varphi _2(x)\phi _2(y),
\varphi _i(x)>0, \phi
_i(y)>0, 0<x,y<1, i=1,2$$ are continuous functions on $[0,1]$,
$n$ is a positive integer. We prove the following results: when
$n= 1$, either there does not exist, or there exist infinitely
many positive solutions in $C[0,1]$; when $n\geq 2$, there exist
at least $1$, at most $n+1$ positive solutions in $C[0,1]$. Especially,
when $n$ is an odd number greater than $2$, there exist at least $1$, at most
$n$ positive solutions in $C[0,1]$.
Necessary and sufficient conditions are derived for the cases: 1) {\small $n= 1$}, there
exist positive solutions in $C[0,1]$; 2) {\small $n\geq 2$}, there exist
exactly $m(m\in \{1,2,\cdots,n+1\})$
positive solutions in $C[0,1]$. Our results generalize the existing results in the literature, and
their usefulness is shown by examples presented in this paper.

In essence, the number of positive solutions to $(2)$ can be transformed into
determination of real roots of a certain polynomial, which is a century-long, albeit
still active research area in mathematics. The classical Sturm method or Newton formula
can be employed to determine the real root distribution of polynomials [4-7],
but the Sturm method is inefficient in establishing discriminant systems for
high-order polynomials with symbolic coefficients [4,6,7], and Newton formula
involves in recursive procedure to determine the real roots, thus it is difficult to
establish explicit criteria [4,5,7].

More recently, Yang et al. established the complete discrimination system for polynomials,
which can give a set of explicit expressions based on the coefficients of polynomials
to determine the root distribution of polynomials [6,7].

Let
\begin{center}
$f(x)=a_0x^n+a_1x^{n-1}+\cdots +a_n\in P^n,$
\end{center}
the Sylvester matrix of $f(x)$ and its derivative $f^{^{\prime }}(x)$ [6,7]
\begin{center}
$\left[
\begin{array}{ccccccccc}
a_0 & a_1 & a_2 & \cdots  & a_{n-1} & a_n &  &  &  \\
0 & na_0 & (n-1)a_1 & \cdots  & 2a_{n-2} & a_{n-1} &  &  &  \\
& a_0 & a_1 & \cdots  & a_{n-2} & a_{n-1} & a_n &  &  \\
& 0 & na_0 & \cdots & 3a_{n-3} & 2a_{n-2} & a_{n-1} &  &  \\
&  &  & \cdots  & \cdots & \cdots  &  &  &  \\
&  &  & \cdots  & \cdots & \cdots  &  &  &  \\
&  &  &  & a_0 & a_1 & a_2 & \cdots  & a_n \\
&  &  &  & 0 & na_0 & (n-1)a_1 & \cdots  & a_{n-1}
\end{array}
\right] $
\end{center}
is called the discrimination matrix of $f(x)$, denoted as $Discr(f).$

\begin{center}
$[D_1(f),D_2(f),\cdots ,D_n(f)]$
\end{center}
the even-order principal minor sequence of $Discr(f)$, is called the
discriminant sequence of $f(x)$.

\begin{center}
$[sign(D_1),sign(D_2),\cdots ,sign(D_n)]$
\end{center}
is called the sign list of the discriminant sequence $[D_1,D_2,\cdots ,D_n]$,
where $sign(\cdot)$ is the sign function, i.e.,
$$
sign(x)=\left\{
\begin{array}{ll}
1\quad\quad\quad &\mbox{if } x>0,\\
0&\mbox{if } x=0,\\
-1&\mbox{if } x<0.\\
\end{array}
\right .
$$%

Given a sign list $[s_1,s_2,\cdots ,s_n],$ we can construct a revised sign list
\begin{center}
$[\varepsilon _1,\varepsilon _2,\cdots ,\varepsilon _n]$
\end{center}%
as follows:%

$1)$If $[s_i,s_{i+1},\cdots ,s_{i+j}]$ is a section of the given sign list
and
$s_i\neq 0;s_{i+1}=s_{i+2}=\cdots =s_{i+j-1}=0;s_{i+j}\neq 0,$ then replace
the subsection consisting of all $0$ elements
$$[s_{i+1},s_{i+2},\cdots ,s_{i+j-1}]$$
by the following subsection with equal number of terms
$$[-s_i,-s_i,s_i,s_i,-s_i,-s_i,s_i,s_i,-s_i,\cdots ]$$
i.e.,$\varepsilon _{i+r}=(-1)^{\displaystyle[\frac {r+1}{2}]}\cdot s_i,
r=1,2,\cdots ,j-1.$%

$2)$Let $\varepsilon _k=s_k$ for all other terms, i.e., all other terms
remain the same.

{\bf Lemma 1} [6,7]\ \
Given the polynomial with real coefficients $f(x)=a_0x^n+a_1x^{n-1}+\cdots +a_n\in P^n.$
If the number of sign changes in the revised sign list of its discriminant sequence
is $\nu ,$ and the number of non-zero elements in the revised sign list is $\mu,$
then the number of distinct real roots of $f(x)$ is $\mu-2\nu .$

{\bf Remark 1}\ \
The discriminant sequence of $f(x)$ can also be constructed by the principal minors of
the Bezout matrix
of $f(x)$ and $f^{^{\prime }}(x)$ [6,7]; the number of distinct real roots of $f(x)$
can also be determined by the sign difference of Bezout matrix of $f(x)$
and $f^{^{\prime }}(x)$ [6,7].

{\bf Remark 2}\ \
The complete discrimination system for polynomials can also be used to determine the number
and the multiplicity of complex roots [6,7].

Yang and Xia also proposed a method to determine the number of positive (negative) roots
of a polynomial [8], which is similar to Lemma 1 in principle, but is more efficient.

{\bf Lemma 2} [8]\ \
Given the polynomial with real coefficients $f(x)=a_0x^n+a_1x^{n-1}+\cdots +a_n\in P^n,$
$a_0\neq 0,a_n\neq 0.$ Let $h(x)=f(-x)$ and $\{d_1,d_2,\cdots,d_{2n+1}\}$ be the sequence of
the principal minors of the discrimination matrix $Discr(h)$ of $h(x)$.
If the number of sign changes in the revised sign list of the sequence
$\{d_1d_2,d_2d_3,\cdots,d_{2n}d_{2n+1}\}$ is $\nu ,$ and the number
of non-zero elements in the revised sign list is $\mu,$
then the number of distinct positive roots of $f(x)$ is $\mu-2\nu .$

\vskip 0.7cm
\section{ Main Results}

\vskip 0.5cm
Consider the problem of determining the number of positive solutions in $C[0,1]$
to the integral equation of the following form
\begin{equation}
\varphi (x)=\int_0^1k(x,y)\varphi ^n(y)dy
\end{equation}
where
$$k(x,y)=\varphi _1(x)\phi _1(y)+\varphi _2(x)\phi _2(y),
\varphi _i(x)>0,\phi
_i(y)>0,0<x,y<1,i=1,2$$
are continuous functions on $[0,1]$,$n$ is a positive integer.

Denote
$$a_{n-i,i}=
C_n^i\int_0^1\phi _1(y)\varphi _1^{n-i}(y)\varphi
_2^i(y)dy,\ \
i=0,1,\cdots,n ,$$
$$b_{n-i,i}=
C_n^i\int_0^1\phi _2(y)\varphi _1^{n-i}(y)\varphi
_2^i(y)dy,\ \
i=0,1,\cdots,n,$$
$$\alpha _i=b_{n-i,i}-a_{n-i+1,i-1},\ i=1,2,\cdots,n,\ \
\alpha _0=b_{n,0},\ \ \alpha _{n+1}=-a_{0,n},\ \ $$
where $C_n^i,i=0,1,\cdots,n,$ stand for the combinatorial number. Our main result
is as follows:

{\bf Theorem 1}\ \
When $n=1$, either there does not exist, or there exist infinitely many
positive solutions in $C[0,1]$ to the integral equation (3). The necessary and sufficient
conditions for the existence of positive solutions in $C[0,1]$ are
$a_{1,0}-1<0$ and
$(a_{1,0}-1)(b_{0,1}-1)-a_{0,1}b_{1,0}=0.$

{\bf Theorem 2}\ \
When $n\geq 2$, there exist at least $1$, at most $n+1$ positive solutions in $C[0,1]$
to the integral equation (3). Especially,
when $n$ is an odd number greater than $2$, there exist at least $1$, at most
$n$ positive solutions in $C[0,1]$.

{\bf Theorem 3}\ \
When $n\geq 2$, the necessary and sufficient
conditions for the existence of exactly $m (m\in\{1,2,\cdots,n+1\})$
positive solutions in $C[0,1]$ to the integral equation (3) are:
the number of sign changes $\nu$ in the revised sign list of the discriminant
sequence of the polynomial $f(s):=\sum\limits_{i=0}^{n+1}\alpha_i s^{2(n+1-i)}$
and the number of its non-zero elements $\mu$ satisfy $m=\frac {\mu-2\nu}{2}$;
or, equivalently, the number of sign changes $\nu$ in the revised sign list of the sequence
$\{d_1d_2,d_2d_3,\cdots,d_{2n+2}d_{2n+3}\}$ and the number of its non-zero
elements $\mu$ satisfy $m=\frac {\mu -2\nu}{2},$ where $\{d_1,d_2,\cdots,d_{2n+3}\}$
is the sequence of the principal minors of the discriminant matrix $Discr(h)$ of
$h(s):=\sum\limits_{i=0}^{n+1}\alpha_i (-s)^{n+1-i}$.

Specifically, when $n=2$, denote
$$p=\frac {\alpha _1}{\alpha _0},r=\frac
{\alpha _2}{\alpha _0},t=\frac {\alpha _3}{\alpha _0}<0,$$
$$\Delta _1=p^2-3r,\Delta _2=rp^2+3tp-4r^2,$$
$$\Delta _3=-4r^3+18rtp+p^2r^2-4p^3t-27t^2,$$
\begin{equation}
[D_1,D_2,D_3,D_4,D_5,D_6]=[1,-p,-p\Delta _1,\Delta _1\Delta _2,\Delta _2\Delta
_3,-t\Delta _3^2].
\end{equation}
then we have

{\bf Corollary 1}\ \
There exist at least $1$, at most $3$ positive solutions in $C[0,1]$ to the integral
equation (1).

{\bf Corollary 2}\ \
The necessary and sufficient conditions for the integral
equation (1) to have exactly $3$ positive solutions in $C[0,1]$ are
$p<0,\Delta _1>0,\Delta _2>0,\Delta _3>0.$

{\bf Corollary 3}\ \
The necessary and sufficient conditions for the integral
equation (1) to have exactly $2$ positive solutions in $C[0,1]$ are
$p<0,\Delta _1>0,\Delta _2>0,\Delta _3=0.$

{\bf Corollary 4}\ \
The necessary and sufficient conditions for the integral
equation (1) to have exactly $1$ positive solutions in $C[0,1]$ are
$p\geq 0,$ or $\Delta _1\leq 0,$ or $\Delta _2\leq 0,$ or $\Delta _3<0.$

{\bf Remark 3}\ \
If $n$ is even, the integral equation (3) does not have any negative solutions
in $C[0,1]$.

{\bf Remark 4}\ \
If $n$ is odd, since $\varphi(x)$ is a positive solution in $C[0,1]$ to
the integral equation (3) if and only if $-\varphi(x)$ is a negative solution in
$C[0,1]$ to the integral equation (3), thus, when $n=1$, the integral equation (3)
either does not have, or has infinitely many negative solutions in $C[0,1]$; when
$n$ is odd and greater than $2$, the integral equation (3) has at least $1$,
at most $n$ negative solutions in $C[0,1]$.

{\bf Remark 5}\ \
When $n=1,$ the necessary and sufficient conditions
for existence of negative solutions in $C[0,1]$ to
the integral equation (3) are the same as in
Theorem 1. When
$n$ is odd and greater than $2$, the necessary and sufficient conditions
for existence of exactly $m (m\in\{1,2,\cdots,n\})$ negative solutions in $C[0,1]$
to the integral equation (3) are the same as in Theorem 3.

{\bf Remark 6}\ \
Our method can be extended to the case when the integral kernel $k(x,y)$ is taken as
$\sum\limits_{i=1}^{l}\varphi_i(x)\phi_i(y)$, where $\varphi_i(x)>0, \phi_i(y)>0,$
$0<x,y<1,i=1,2,\cdots,l,$ are continuous functions on $[0,1]$.

{\bf Remark 7}\ \
The conclusions in [3] are equivalent to Corollaries 1,2,3 above.

\vskip 0.7cm
\section{ Proof of the Theorems}

\vskip 0.5cm
{\bf Proof of Theorem 1}\ \
When $n=1$, the integral equation (3) becomes
\begin{equation}
\label{(5)}
\varphi (x)=\int_0^1k(x,y)\varphi (y)dy
\end{equation}
Thus, we have
$$\varphi (x)=\varphi _1(x)\int_0^1\phi _1(y)\varphi (y)dy+\varphi
_2(x)\int_0^1\phi _2(y)\varphi (y)dy$$
If $\varphi (x)$ is a positive solution in $C[0,1]$ to equation (5), then
$\varphi (x)$ can be expressed as
$\varphi (x)=\lambda _1\varphi _1(x)+\lambda _2\varphi _2(x),$
where $\lambda _1>0,\lambda _2>0$ are coefficients to be determined.
Taking it into equation (5), we get the following system of algebraic equations
\begin{equation}
\label{(6)}
\left\{
\begin{array}{c}
a_{1,0}\lambda _1+a_{0,1}\lambda _2=\lambda _1 \\
b_{1,0}\lambda _1+b_{0,1}\lambda _2=\lambda _2
\end{array}
\right.
\end{equation}
where
$a_{1,0}=\int_0^1\phi _1(y)\varphi _1(y)dy,\ $
$a_{0,1}=\int_0^1\phi _1(y)\varphi _2(y)dy,\ $
$b_{1,0}=\int_0^1\phi _2(y)\varphi _1(y)dy,\ $
$b_{0,1}=\int_0^1\phi _2(y)\varphi _2(y)dy.$
Apparently, the necessary and sufficient conditions for the system of algebraic
equations (6) to have positive solutions
$\lambda _1,\lambda _2$ are
$a_{1,0}-1<0$, and
$(a_{1,0}-1)(b_{0,1}-1)-a_{0,1}b_{1,0}=0.$
Moreover, if $\varphi (x)$ is a positive solution to equation (5), then, obviously,
for any positive constant number
$c, c\varphi (x)$ is also a positive solution to equation (5). Thus,
there are infinitely many positive solutions in $C[0,1]$ to equation (5).
This completes the proof.

{\bf Lemma 3}\ \
The system of equations
\begin{equation}
\label{(7)}
\left\{
\begin{array}{c}
a_{n,0}x^n+a_{n-1,1}x^{n-1}y+a_{n-2,2}x^{n-2}y^2+\cdots
+a_{1,n-1}xy^{n-1}+a_{0,n}y^n=x \\
b_{n,0}x^n+b_{n-1,1}x^{n-1}y+b_{n-2,2}x^{n-2}y^2+\cdots
+b_{1,n-1}xy^{n-1}+b_{0,n}y^n=y
\end{array}
\right.
\end{equation}
$$a_{n-i,i}>0,\ \ b_{n-i,i}>0,\ \ i=0,1,2,\cdots ,n.$$
has at least $1$, at most $n+1$ (at most $n$, when $n$ is odd) positive solutions,
where $n\geq 2.$

{ Proof}\ \
Let
$$p(x,y)=\frac x{a_{n,0}x^n+a_{n-1,1}x^{n-1}y+\cdots
+a_{0,n}y^n},\ \ q(x,y)=\frac y{b_{n,0}x^n+b_{n-1,1}x^{n-1}y+\cdots
+b_{0,n}y^n},$$
$$ x>0,\ y>0,$$
then
$$p(\kappa x,\kappa y)=\frac 1{\kappa ^{n-1}}p(x,y),\ \ q(\kappa x,\kappa
y)=\frac 1{\kappa ^{n-1}}q(x,y),\ \kappa >0.$$
Let
$$E=\{x|p(x,1)=q(x,1)\},$$
then the number of positive solutions to the system of equations (7) is
equal to the number of elements in $E$. In fact, if $(x,y)$ is a positive solution
to (7), then
$$ p(x,y)=q(x,y)=1,$$
thus $\frac {x}{y}\in E.$ Conversely, if $x\in E,$ since $n\geq 2,$ it is easy to
verify that
$(\sqrt[n-1]{p(x,1)}x,\sqrt[n-1]{p(x,1)})$ is a positive solution to (7).%

\noindent Suppose $x\in E,$ by
$p(x,1)=q(x,1)$, we have
$$
b_{n,0}x^{n+1}+(b_{n-1,1}-a_{n,0})x^n+(b_{n-2,2}-a_{n-1,1})x^{n-1}+\cdots
+(b_{0,n}-a_{1,n-1})x-a_{0,n}=0
$$
Namely
\begin{equation}
\label{(8)}
\alpha _0x^{n+1}+\alpha _1x^n+\alpha _2x^{n-1}+\cdots +\alpha _nx+\alpha
_{n+1}=0
\end{equation}
where
$$\alpha _i=b_{n-i,i}-a_{n-i+1,i-1},\ i=1,2,\cdots,n,\ \
\alpha _0=b_{n,0},\ \ \alpha _{n+1}=-a_{0,n},\ \ $$
Since $\alpha _0=b_{n,0}>0,\alpha _{n+1}=-a_{0,n}<0,$
equation (8) has at least $1$, at most $n+1$ positive roots. Especially, when
$n>2$ and is odd, since equation (8) has at least $1$ negative root, it
has at most $n$ positive roots. This completes the proof.

{\bf Proof of Theorems 2 and 3}\ \
When $n\geq 2$, since
$$
\begin{array}{ll}
\varphi (x) & =\int_0^1k(x,y)\varphi ^n(y)dy\\
& =\varphi _1(x)\int_0^1\phi _1(y)\varphi ^n(y)dy+\varphi
_2(x)\int_0^1\phi _2(y)\varphi ^n(y)dy\\
\end{array}
$$
similar to the proof of Theorem 1, the positive solution $\varphi (x)$ in
$C[0,1]$ to the integral equation (3) can be expressed as
$\varphi (x)=\lambda _1\varphi _1(x)+\lambda _2\varphi _2(x),$
where $\lambda _1>0,\lambda _2>0$ are coefficients to be determined.
Taking it into equation (3), by a simple but lengthy calculation,
we see that  $\lambda _1, \lambda _2$  should be positive solutions to
the following system of algebraic equations
\begin{equation}
\label{(9)}
\left\{
\begin{array}{c}
a_{n,0}\lambda _1^n+a_{n-1,1}\lambda _1^{n-1}\lambda _2
+a_{n-2,2}\lambda _1^{n-2}\lambda _2^2+\cdots
+a_{1,n-1}\lambda _1\lambda _2^{n-1}+a_{0,n}\lambda _2^n=\lambda _1 \\
b_{n,0}\lambda _1^n+b_{n-1,1}\lambda _1^{n-1}\lambda _2
+b_{n-2,2}\lambda _1^{n-2}\lambda _2^2+\cdots
+b_{1,n-1}\lambda _1\lambda _2^{n-1}+b_{0,n}\lambda _2^n=\lambda _2
\end{array}
\right.
\end{equation}
where
$$a_{n-i,i}=
C_n^i\int_0^1\phi _1(y)\varphi _1^{n-i}(y)\varphi
_2^i(y)dy,\ \
i=0,1,\cdots,n ,$$
$$b_{n-i,i}=
C_n^i\int_0^1\phi _2(y)\varphi _1^{n-i}(y)\varphi
_2^i(y)dy,\ \
i=0,1,\cdots,n.$$
By Lemma 3, we complete
the proof of Theorem 2.

\noindent Moreover, from the proof of Lemma 3, we know that finding the positive
solutions to the system of algebraic equations (9) or (7) can be transformed into
finding the positive solutions to equation (8). Applying Lemmas 1 and 2 to equation (8),
we complete the proof of Theorem 3.

{\bf Proof of Corollaries 1,2,3,4}\ \
Some notations in this proof are defined in Section 2.

\noindent Corollary 1 is a direct consequence of Theorem 2.

\noindent When $n=2$, equation (8) becomes
\begin{equation}
\label{(10)}
\alpha _0x^{3}+\alpha _1x^2+\alpha _2x+\alpha_{3}=0
\end{equation}
By a direct computation, we know that the discriminant sequence
$[D_1,D_2,D_3,D_4,D_5,D_6]$ of the polynomial
$f(s):= \alpha _0s^{6}+\alpha _1s^4+\alpha _2s^2+\alpha_{3}$ is determined by (4)
(up to a positive factor).

\noindent Since $t<0,$ it is easy to see that the number of sign changes $\nu $
in the revised sign list of $[D_1,D_2,D_3,D_4,D_5,D_6]$ and the number of its
non-zero elements $\mu$ satisfy $6=\mu-2\nu$ if and only if the revised sign list
of $[D_1,D_2,D_3,D_4,D_5,D_6]$ is $[1,1,1,1,1,1],$ which is equivalent to
$p<0,\Delta _1>0,\Delta _2>0,\Delta _3>0.$ This completes the proof of Corollary 2.

\noindent Similarly, the number of sign changes $\nu $
in the revised sign list of $[D_1,D_2,D_3,D_4,D_5,D_6]$ and the number of its
non-zero elements $\mu$ satisfy $4=\mu-2\nu$ if and only if the revised sign list
of $[D_1,D_2,D_3,D_4,D_5,D_6]$ is $[1,1,1,1,0,0],$ which is equivalent to
$p<0,\Delta _1>0,\Delta _2>0,\Delta _3=0.$ This completes the proof of Corollary 3.

\noindent Combining Corollaries 1,2,3, we get Corollary 4.

\vskip 0.7cm
\section{ Some Illustrative Examples}

\vskip 0.5cm
{\bf Example 1}\ \
Consider the integral equation
\begin{equation}
\label{(11)}
\varphi (x)=\int_0^1(\frac{6}{5}xy+\frac{3}{5}y)\varphi (y)dy,\ \ 0\leq x\leq 1
\end{equation}
Let
$$\varphi _1(x)=\frac{6}{5}x,\ \phi _1(y)=y,\
\varphi _2(x)=\frac{3}{5},\ \phi _2(y)=y.$$
Then, it is easy to get
$a_{1,0}=b_{1,0}=\frac{2}{5},\  a_{0,1}=b_{0,1}=\frac{3}{5}.$
The conditions in Theorem 1 are met. Hence, there are infinitely many
positive solutions in $C[0,1]$. In fact,
$\varphi (x)=c(\frac{6}{5}x+\frac{3}{5}),\forall c>0$ are such solutions.

{\bf Remark 8}\ \
From the proof of theorems and the example above, we can see that, for a
given integral equation,
we can not only determine the number of its positive solutions, but also
find the positive solutions explicitly by solving the algebraic equation (8).

{\bf Example 2}\ \
Consider the integral equation
\begin{equation}
\label{(12)}
\begin{array}{ll}
\varphi (x)= & \int_0^1[18\max \{\varepsilon ,-2x+1+\varepsilon \}
+\max \{\frac{\varepsilon }{3},\frac{1}{3}(2x-1+\varepsilon)\}\\
& \times \max \{6,272y-130\}]\varphi ^n(y)dy,\ \ 0\leq x\leq 1,\\
\end{array}
\end{equation}
where $\varepsilon \geq 0,n=1$ or $2.$

\noindent Let
$$\varphi _1(x)=\max \{\varepsilon ,-2x+1+\varepsilon \},\ \phi _1(y)=18,$$
$$\varphi _{2}(x)=\max \{\frac{\varepsilon }{3},
\frac{1}{3}(2x-1+\varepsilon
)\},\ \phi _{2}(y)=\max \{6,272y-130\}.$$

\noindent When $n=1$, using the notations in Section 2 and by a simple computation, we can get
$$a_{1,0}=\int_0^1\phi _1(y)\varphi _1(y)dy = 18\varepsilon + \frac 92,\
a_{0,1}=\int_0^1\phi _1(y)\varphi _2(y)dy = 6\varepsilon +\frac 32;$$
$$b_{1,0}=\int_0^1\phi _2(y)\varphi _1(y)dy = 40\varepsilon +\frac 32,\
b_{0,1}=\int_0^1\phi _2(y)\varphi _2(y)dy =
\frac{40}3\varepsilon +\frac{145}{18}.$$
Since $a_{1,0}-1>0,$ the conditions in Theorem 1 are not met. Thus, equation (12)
does not have any positive solutions in $C[0,1]$.

\noindent Similarly, when $n=2$, by a simple computation, we can get
$$a_{2,0} = 3+9\varepsilon +18\varepsilon ^2,\
a_{1,1} = 6\varepsilon +12\varepsilon ^2,\
a_{0,2} = \frac 13+\varepsilon +2\varepsilon ^2;$$
$$b_{2,0}= 1+3\varepsilon +40\varepsilon ^2,\
b_{1,1}= \frac{154}9\varepsilon +\frac{80}3\varepsilon ^2,\
b_{0,2}= 2+\frac{145}{27}\varepsilon +\frac{40}9\varepsilon ^2;$$
$$\alpha _{0}=b_{2,0}=1+3\varepsilon +40\varepsilon ^{2},\
\alpha _{1}=b_{1,1}-a_{2,0}
= \frac{26}{3}\varepsilon ^{2}+\frac{73}{9}\varepsilon -3,\ $$
$$\alpha _{2}=b_{0,2}-a_{1,1}=-\frac{68}{9}%
\varepsilon ^{2}-\frac{17}{27}\varepsilon+2,\
\alpha _3=-a_{0,2}= -2\varepsilon ^2-\varepsilon-\frac 13.\ $$

\noindent Let
$$p=\frac{\alpha _1}{\alpha _0},\
r=\frac{\alpha _{2}}{\alpha _{0}},\
t=\frac{\alpha _{3}}{\alpha _{0}};$$
$$\Delta _1=p^2-3r,\ \Delta _2=rp^2+3tp-4r^2,\ $$
$$\Delta _3=-4r^3+18rtp+p^2r^2-4p^3t-27t^2,\ $$
we have (up to a positive factor)
$$p=-1+2.7037\varepsilon +2.8889\varepsilon ^2,$$
$$\Delta _1=-65.959\varepsilon ^2+94.716\varepsilon ^3-
21.593\varepsilon +327.26\varepsilon ^4+1,$$
$$\Delta _2=702.22\varepsilon ^4-1291.9\varepsilon ^5+
255.78\varepsilon ^3-2356.3\varepsilon ^6
-78.517\varepsilon ^2-26.207\varepsilon +1,$$
$$\Delta _3=1-27.778\varepsilon -1.4371\times 10^5\varepsilon
^6-23275.0\varepsilon ^4-1.0374\times 10^5\varepsilon ^5
-63.724\varepsilon ^2-1222.6\varepsilon ^3.$$
By numerical computations, it is easy to get
\begin{center}
The real roots of $p=0$ are $-1.2197$ and $0.2838;$

The real roots of $\Delta _1=0$ are $0.041426$ and $0.45024;$

The real roots of $\Delta _2=0$ are $-0.70495$ and $0.034952;$

The real roots of $\Delta _3=0$ are $-0.21287$ and $0.03143.$
\end{center}

\noindent Hence, by Corollaries 2,3,4, it is easy to know that there exists a
positive number $r_0 \approx 0.03143$ (note here the difference between the
exactness of the conditions in Corollaries 2,3,4 and the inexactness of the
numerical computations above), such that: when
$0\leq \varepsilon <r_0$, equation (12) has 3 positive solutions in $C[0,1]$;
when $\varepsilon = r_0$, equation (12) has 2 positive solutions in $C[0,1]$;
when $\varepsilon >r_0$, equation (12) has 1 positive solutions in $C[0,1]$.

{\bf Remark 9}\ \
The case when $n=2$ in the example above has also been studied in [3].
Our result here is completely consistent with the result in [3].

{\bf Example 3}\ \
Consider the integral equation (12) in the example above. When $n=3, \varepsilon$
is $2$ or $0.2$, determine the number of its positive solutions in $C[0,1]$.

\noindent Similar to Example 2, when $n=3$, using the notations in Section 2 and by a simple computation, we can get
$$a_{3,0}= \frac 94+9\varepsilon +\frac{27}2\varepsilon ^2+18\varepsilon
^3,\
a_{2,1}= 3\varepsilon +\frac{27}2\varepsilon ^2+18\varepsilon ^3,\ $$
$$a_{1,2}=\frac 92\varepsilon ^2+6\varepsilon
^3+\varepsilon ,\
a_{0,3}= \frac 23\varepsilon ^3+\frac 1{12}+\frac 13\varepsilon +\frac
12\varepsilon ^2;$$
$$b_{3,0}= \frac 34+3\varepsilon +\frac 92\varepsilon ^2+40\varepsilon ^3,\
b_{2,1}= \varepsilon +\frac{163}6\varepsilon ^2+40\varepsilon ^3,\ $$
$$b_{1,2}=\frac{299}{%
18}\varepsilon ^2+\frac{40}3\varepsilon ^3+6\varepsilon ,\
b_{0,3}= \frac{163}{54}\varepsilon ^2+\frac{287}{540}+2\varepsilon +%
\frac{37}{27}\varepsilon ^3.$$

$$\alpha _0=\frac 34+3\varepsilon +\frac 92\varepsilon
^2+40\varepsilon ^3,\
\alpha _1=-8\varepsilon +\frac{41}3\varepsilon ^2+22\varepsilon
^3-\frac 94,\ $$
$$\alpha _2=\frac{28}9\varepsilon ^2-\frac{14}3\varepsilon
^3+3\varepsilon ,\
\alpha _3=-\frac{40}{27}\varepsilon ^2+\frac{287}{540}%
+\varepsilon -\frac{125}{27}\varepsilon ^3,\ $$
$$\alpha _4=-\frac 23\varepsilon ^3-\frac 1{12}-\frac
13\varepsilon -\frac 12\varepsilon ^2.$$

\noindent Hence, when
$\varepsilon =2$, we have
$\alpha _0s^8+\alpha _1s^6+\alpha _2s^4+\alpha _3s^2+\alpha _4=
344.75s^8+212.42s^6-18.889s^4-40.431s^2-
8.0833.$ By a simple computation, the revised sign list of its discriminant
sequence is $$[1,-1,-1,-1,1,1,1,-1].$$ By Theorem 3, equation (12) has only 1
positive solution in $C[0,1]$.

\noindent When
$\varepsilon =0.2$, we have
$\alpha _0s^8+\alpha _1s^6+\alpha _2s^4+\alpha _3s^2+\alpha _4=
1.85s^8-3.1273s^6+0.68711s^4+0.63519s^2-0.17533.$ By a simple computation,
the revised sign list of its discriminant sequence is
$$[1,1,1,-1,-1,-1,-1,-1].$$ By Theorem 3, equation (12) has 3
positive solutions in $C[0,1]$.

\vspace*{1.5\baselineskip}
\vskip 1cm
\centerline{\LARGE References}
\vspace*{1.0\baselineskip}

\def\toto#1#2{\centerline{\hbox to0.6cm{#1\hss}
\parbox[t]{14cm}{#2}}\vspace{0.5\baselineskip}}
{\parindent=0pt

\toto{1}
{C. Corduneanu, Integral Equations and Stability of Feedback Systems,
{ Springer-Verlag, London and New York,} 1973.}
\toto{2}
{R. Courant and D. Hilbert, Methods of Mathematical Physics,
{ Interscience Publishing Company, New York,} 1953.}
\toto{3}
{P. Yao, On the number of positive solutions to an integral equation with
rank 2 kernel,
{ Journal of Mathematical Physics,} 1991,{\bf 11:} 274-279.}
\toto{4}
{F. R. Gantmacher, The Theory of Matrices,
{ Chelsea, New York,} 1960.}
\toto{5}
{W. H. Greub, Linear Algebra,
{ Springer-Verlag, Berlin,} 1967.}
\toto{6}
{L. Yang, X. R. Hou and Z. B. Zeng,
A complete discrimination system for polynomials.
{\it Science in China,} {\bf E-39:} 628-646, 1996
}
\toto{7}
{L. Yang, J. Z. Zhang and X. R. Hou, Nonlinear Algebraic Equations
and Machine Proving(Nonlinear Science Series),
{ Shanghai Science and Education Press, Shanghai,} 1996.}
\toto{8}
{L. Yang and B. C. Xia,
Explicit criterion to determine the number of positive
roots of a polynomial.
{\it Mathematics-Mechanization Research Preprints,} {\bf 15:} 134-145, 1997.}
}

\end{document}